\documentclass[preprintnumbers,article,amsmath,amssymb,floatfix,10pt,prd,onecolumn,
superscriptaddress,nofootinbib]{revtex4-2}
\usepackage{bm}
\usepackage{amsfonts}
\usepackage{latexsym}
\usepackage{graphicx}
\usepackage{amsmath}
\usepackage{dirtytalk}
\usepackage{palatino}
\usepackage{mathpazo}
\usepackage{textcomp}
\usepackage{float}
\usepackage{enumerate}
\usepackage{booktabs}
\usepackage{dcolumn}
\usepackage{ragged2e}
\usepackage{hyperref}
\hypersetup{colorlinks,citecolor=blue}
\hypersetup{colorlinks=true,linkcolor=red,filecolor=magenta,    urlcolor=blue}
\usepackage{amsmath}
\usepackage{xcolor}
\usepackage{orcidlink}
\usepackage{epsfig}
\usepackage{caption}
\usepackage{subcaption}
\usepackage{commath}
\usepackage[displaymath,mathlines,pagewise]{lineno}
\def\jnl@style{\it}
\def\aaref@jnl#1{{\jnl@style#1}}

\def\aaref@jnl#1{{\jnl@style#1}}

\def\aj{\aaref@jnl{AJ}}                   
\def\apj{\aaref@jnl{ApJ}}                 
\def\apjl{\aaref@jnl{ApJ}}                
\def\apjs{\aaref@jnl{ApJS}}               
\def\apss{\aaref@jnl{Ap\&SS}}             
\def\aap{\aaref@jnl{A\&A}}                
\def\aapr{\aaref@jnl{A\&A~Rev.}}          
\def\aaps{\aaref@jnl{A\&AS}}              
\def\mnras{\aaref@jnl{Mon.~Not.~Roy.~Astron.~Soc.}}             
\def\prd{\aaref@jnl{Phys.~Rev.~D}}        
\def\prc{\aaref@jnl{Phys.~Rev.~C}}  
\def\prl{\aaref@jnl{Phys.~Rev.~Lett.}}    
\def\qjras{\aaref@jnl{QJRAS}}             
\def\skytel{\aaref@jnl{S\&T}}             
\def\ssr{\aaref@jnl{Space~Sci.~Rev.}}     
\def\zap{\aaref@jnl{ZAp}}                 
\def\nat{\aaref@jnl{Nature}}              
\def\aplett{\aaref@jnl{Astrophys.~Lett.}} 
\def\apspr{\aaref@jnl{Astrophys.~Space~Phys.~Res.}} 
\def\physrep{\aaref@jnl{Phys.~Rep.}}      
\def\physscr{\aaref@jnl{Phys.~Scr}}       
\def\commat{\aaref@jnl{Comm.~Math.~Phys.}}              
\def\science{\aaref@jnl{Science}}               
\def\cqg{\aaref@jnl{Classical Quant.~Grav.}}            
\def\jpcs{\aaref@jnl{JPCS}}                                     
\def\ijmpd{\aaref@jnl{Int.~J.~Mod.~Phys.~D}}                    
\def\grg{\aaref@jnl{Gen.~Relat.~Gravit.}}               
\def\rpp{\aaref@jnl{Rep.~Prog.~Phys.}}          
\def\npa{\aaref@jnl{Nucl.~Phys.~A}}        
\def\lrr{\aaref@jnl{Living Rev.~Rel.}}                   
\def\jcap{\aaref@jnl{J.~Cosmology Astropart.~Phys.}}    
\def\rmp{\aaref@jnl{Rev.~Mod.~Phys.}}   
\def\epjc{\aaref@jnl{Eur.~Phys.~J.~C}} 
\def\plb{\aaref@jnl{~Phy.~Lett.~B}} 
\def\mpla{\aaref@jnl{Mod.~Phy.~Lett.~A}} 
\def\arxiv{\aaref@jnl{arxiv.org}}

\allowdisplaybreaks[1]

\begin{document}
\color{black}       
%
\title{Cosmic evolution of the Kantowski-Sachs universe in the context of a bulk viscous string in teleparallel gravity}

\author{S.R.Bhoyar \orcidlink{0000-0001-8427-4540}}
\email{drsrb2014@gmail.com}
\affiliation{Department of Mathematics, Phulsing Naik Mahavidyalaya Pusad-445216 Dist. Yavatmal (India)}

\author{Yash B. Ingole \orcidlink{0009-0006-7208-1999}}
\email{ingoleyash01@gmail.com}
\affiliation{Department of Mathematics, Phulsing Naik Mahavidyalaya Pusad-445216 Dist. Yavatmal (India)}
%

\begin{abstract}
In the present work, we analyzed the Kantowski-Sachs cosmological model and teleparallel gravity, where a bulk viscous fluid containing one-dimensional cosmic strings is the source for the energy$-$momentum tensor. To obtain the deterministic solution of the field equations, we employed the proportionality condition linking the shear scalar $(\sigma)$ and the expansion scalar $(\theta)$, establishing a relationship between metric potentials. Another approach employed is the hybrid expansion law (HEL). The discussion focuses on the behavior of the accelerating universe concerning the specific choice of a nonlinear (or power law model) of teleparallel gravity $f(T)=\alpha T + \beta T^m$, where $T$ is the torsion scalar, $\alpha$ , and $\beta$  are model parameters and $m$ is restricted to greater than or equal to 2. The effective equation of the state parameter $(\omega_{eff})$  of models will support the acceleration of the universe. We observed that the null and weak energy conditions are obeyed but violate the strong energy condition as per the present accelerating expansion. Under specific model parameter constraints, the universe shows a transition from a decelerating to an accelerating phase.

\textbf{Keywords:} Kantowski$-$Sachs universe; Bulk viscous string; Hybrid expansion law; $f(T)$ gravity.

\end{abstract}

\maketitle

\date{\today}

\section{Introduction}
Numerous cosmological studies over the last two decades have come together to explain that our universe is presently undergoing an accelerated expansion phase, as indicated by data gathered from Type Ia supernovae \cite{riess1998observational,R2}, baryon acoustic oscillations (BAO) \cite{R3,R4}, large-scale structures \cite{R6,R7}, galaxy redshift surveys \cite{R5}, and cosmic microwave background radiations (CMBR) \cite{R8,R9}. In addition, the most important components driving the accelerated expansion of the universe are dark energy (DE) and dark matter (DM) which contribute 95\% of the total  energy of the universe  \cite{gadbail2021viscous,mandal2020complete}. The vacuum energy (also known as the cosmological constant, $\Lambda$) with a constant equation of state parameter (EoS) $\omega_\Lambda =-1$ is the simplest and most appealing candidate for DE \cite{R10}. Furthermore, the cosmological constant $(\Lambda)$ has been influenced by two crucial issues: the cosmic coincidence problem and the fine-tuning problem \cite{copeland2006dynamics,solanki2022statefinder}. To solve the aforementioned cosmological issues, time-varying dark energy models, such as the quintessence \cite{carroll1998quintessence,fujii1982origin}, k-essence \cite{chiba2000kinetically,armendariz2000dynamical}, and perfect fluid models (such as the Chaplygin gas model) \cite{bento2002generalized,kamenshchik2001alternative}, have been proposed in the literature. Another approach to clarifying the current acceleration of the universe is to alter (or modify) the geometry of spacetime, which can be accomplished by modifying the left-hand side of Einstein's equation. Modified theories of gravity represent the geometric extension of Einstein's general theory of relativity, in which acceleration of the universe can be accomplished through modifications to the Einstein$-$Hilbert action of general relativity (GR). Recently, cosmologists have shown keen interest in modified theories of gravity for understanding the role of DE. Within modified gravity, the source of DE is acknowledged as a modification of gravity \cite{solanki2021cosmic}.

Some modified theories of gravity include the $f(R)$ theory of gravity, the modification of GR introduced in \cite{buchdahl1970non,saridakis2021modified},
the $f (R, T )$ theory, an extension of $f (R)$ gravity coupled
with the trace of energy-momentum tensor $T$ \cite{harko2011f,shabani2013f,sahoo2018wormholes}, the
$f (G)$ gravity \cite{de2009construction,goheer2009coexistence,bamba2017energy}, $f (R, G)$ theory \cite{elizalde2010lambdacdm,bamba2010finite},
$f(T)$ theory \cite{ferraro2007modified,linder2010einstein,bamba2011equation}, $f(T, B)$ theory \cite{bahamonde2017nonlocal}, $f(Q, T)$ theory \cite{arora2021constraining,xu2019f,solanke2023anisotropic}, and
$f(G)$ theory \cite{nojiri2008inflation}. where $R,T,G,B$, and $Q$ are the Ricci, Torsion, Gauss, Bonnet, and nonmetricity scalars, respectively. To provide a comprehensive framework for studying equivalent representations of gravity within the context of metric$-$affine geometries, the geometric trinity is becoming popular. Geometric trinity consists of three fundamental theories: general relativity (GR), the teleparallel equivalent of general relativity (TEGR), and the symmetric teleparallel equivalent of general relativity (STEGR) \cite{beltran2019geometrical}. These theories represent different formulations of gravity within the framework of metric$-$affine geometries. While GR is based on the metric tensor and curvature, TEGR relies on torsion and tetrads, and STEGR is built upon nonmetricity. Despite their differences, these theories are dynamically equivalent, as demonstrated through variational methods, field equations, and solutions. The geometric trinity of gravity provides insights into the fundamental concepts of gravitational fields and offers a unified view of gravitational interactions, highlighting the importance of understanding the mathematical and physical aspects of each theory \cite{capozziello2022comparing,capozziello2023role}.
In the context of gravitational theories, $f(T)$ gravity and the TEGR are closely related. The gravity $f(T)$  is an extension of the TEGR, where TEGR is a specific case of $f(T)$ gravity. In TEGR, the gravitational dynamics are described regarding torsion, which replaces curvature as the fundamental geometric quantity. On the other hand, $f(T)$ gravity extends the TEGR by considering more general functions of the torsion scalar $T$. By introducing these additional functions, $f(T)$ gravity allows for a broader exploration of gravitational theories beyond the constraints of the TEGR.\cite{capozziello2011extended,aldrovandi2012teleparallel}. The tetrad or vierbein field, an orthonormal basis field in the tangent space is the primary dynamical variable in this theory. $ f(T)$ theories are gaining significant interest \cite{yang2011conformal,wu2010dynamical,bamba2010comment,linder2010einstein} due to their ability to characterize an inflationary expansion without requiring the use of an inflaton field \cite{ferraro2008born,mai2017black,bahamonde2017new}. Any modified theory of gravity will find it appealing that the dynamical equations are always second order as the action of $f(T)$ gravity only contains the first derivatives of the vierbein. In this respect, the $f(T )$ gravity is separated from the metric $f(R)$ gravity,
whose dynamical equations are of order four.

It has been demonstrated that viscosity is one of the most important aspects of exploring the evolution of the universe. According to the current understanding, the expansion of the universe is an outcome of negative pressure in a cosmic fluid \cite{brevik2012turbulence}. In addition, several researchers have claimed that late-time accelerated expansion is influenced by cosmic viscosity. Viscosity theories in cosmology are significant when addressing the early universe i.e., when the temperature was approximately $10^4K$. \cite{arora2020effective}. In cosmic fluid, there are two distinct viscosity coefficients: bulk viscosity $\xi$ and shear viscosity $\eta$. Because of the accepted spatial isotropy of the universe, as described by the Robertson-Walker metric, we exclude shear viscosity. Theoretically, deviations from the local thermodynamic irreversibility of motion cause bulk viscosity to occur. Bulk viscosity occurs in cosmology as a bulk viscous pressure to return the system to thermal equilibrium \cite{okumura2003new}. Due to viscosity, the thermodynamic pressure $p$ is replaced by the bulk viscous pressure $\Bar{p}$, and the bulk viscous pressure is redefined as $\Bar{p}=p-3\xi H$, where $\xi$ is the expected bulk viscous coefficient based on the background geometry and $H$ is the Hubble parameter. Traditional models such as the cosmological constant ($\Lambda$) can successfully describe cosmic acceleration, and the combination of modified gravity and viscosity provides an alternative explanation \cite{bamba2010finite}. This superposition allows for a more comprehensive understanding of the underlying mechanisms driving the accelerated expansion of the universe. This approach opens up possibilities for investigating the interplay between gravitational theories and fluid dynamics in the context of cosmic evolution \cite{bamba2010finite,harko2011f,bhatti2019stability}. Many authors have also considered the theory of bulk viscous fluids as a possible explanation for the acceleration of the expansion of the universe \cite{waga1986bulk,padmanabhan1987viscous,cheng1991bulk,samanta2017imperfect}. Furthermore, Davood \cite{sadatian2019effects} investigated how bulk viscosity affects $f(T)$ gravity. D.C. Maurya et al. \cite{maurya2024modified} studied a dusty string fluid matter source in $f(Q)$ gravity to investigate an anisotropic locally rotational spacetime (LRS) Bianchi type-I model. S. Bekkhozhayev et al. \cite{bekkhozhayev2024impact} explained the accelerated expansion with the help of a bulk viscous fluid coupled with $f(Q,T)$ gravity by assuming the pure dominance of nonrelativistic viscous matter over the universe. Recently, M. Krishna et al. \cite{krishna2024accelerating} used Lyra's geometry in a plane-symmetric metric in the presence of a bulk viscous fluid and one-dimensional strings.  D.D.Pawar et al. D. Pawar et al. \cite{pawar2024perfect} investigated the dynamics of the Bianchi Type-I model in $f(T)$ gravity by determining the presence of a perfect fluid with heat flow in a cosmic medium. G. Gomez et al. proposed the general parametrization of the bulk viscosity \cite{gomez2023new}. R Solanki. et al. \cite{solanki2021cosmic} explained the role of bulk viscosity in the $f(Q)$ gravity framework by assuming a bulk viscous matter-dominated cosmological model with the bulk viscosity coefficient. 

Cosmic strings are topologically stable entities that may manifest during a phase transition in the early universe \cite{kibble1976topology}. Cosmic strings play a pivotal role in the study of the early universe, which emerges during the phase transition after the explosion of the Big Bang as the temperature decreases below a certain critical temperature, as predicted by grand unified theories \cite{zel1974cosmological,kibble1980some,everett1981cosmic,vilenkin1981cosmic}. These cosmic strings possess stress energy and are coupled with the gravitational field; thus, it is interesting to examine the gravitational effects arising from strings. The general relativistic treatment of strings was initiated by Letelier \cite{letelier1979clouds,letelier1983string} and Stachel \cite{stachel1980thickening}. Letelier derived the solution to Einstein's field equations for a cloud of strings with spherical, planar, and cylindrical symmetry. Subsequently, in $1983$, he resolved Einstein's field equations for a cloud of massive strings and acquired cosmological models in Bianchi type-I and Kantowski-Sachs space-times. Bali and Pradhan \cite{bali2007bianchi} investigated string cosmological models of Bianchi type-III with a bulk viscous fluid for massive strings. By considering a cosmic string as a source of the energy-momentum tensor, A.J. Meitei et al. \cite{meitei2024cosmological} examined a string cosmic model in homogeneous but anisotropic Bianchi Type-V space-time in the context of the $f(R, T)$ theory of gravity.  

In this article, we describe the cosmological model with bulk viscosity in the framework of the nonlinear form of teleparallel gravity. Our motivation for this study stems from the abovementioned discussions and analysis. Considering teleparallel gravity with the Kantowski-Sachs universe and bulk viscous fluid is motivated by the need to explore alternative formulations of gravity, understand the role of anisotropies, provide a rich framework for exploring complex interactions between different cosmological factors, and potentially solve outstanding problems in cosmology, such as dark energy, dark matter, and the accelerated expansion of the universe. We consider three models based on the values of model parameters $\alpha$ and $\beta$.  This combination can lead to a better understanding of the universe's large-scale structure and its evolution.

The structure of this paper is as follows: In Sect. \eqref{II} we describe the formulation of $f(T)$ gravity. Section \eqref{III} contains a brief review of the Kantowski-Sachs universe model along with the field equations. In Sect. \eqref{IV}, we describe the field equations and study the three models, which are based on the value of a model parameter. This section also contains a comprehensive examination of physical parameters and kinematic parameters. Section \eqref{V} contains the energy conditions of models I to III, and in Sect. \eqref{VI}, we conclude the paper.

\section{Overview of $f(T)$ gravity}\label{II}

\empty The dynamical object employed in teleparallelism is a vierbein field $e_i(x^\mu)$, where $i=0,1,2,3 $, which represents an orthonormal basis for the tangent space at each point $x^\mu$  of the manifold. This vierbein field satisfies the condition, $e_i.e_j=\eta_{ij}=diag(+1,-1,-1,-1)$. The components of each vector $e_i$  can be expressed as $e_i^{\mu}$,\quad $\mu=0,1,2,3$ on a coordinate basis, denoted as $e_i=e_i^{\mu}\partial_\mu$. It is important to note that Latin indices relate to the tangent space, while Greek indices indicate coordinates on the manifold. The metric tensor $g_{\mu\nu}(x)$ is derived from the dual vierbein by $g_{\mu\nu}(x)=\eta_{ij}e{^i_\mu}(x)e{^j_\nu}(x)$. 

A teleparallel structure which is directly related to the gravitational field generated in space-time by a nontrivial tetrad field. Otherwise, it is possible to define the so-called Weitzenb\"ock connection given as nontrivial tetrad.
\begin{equation}\label{1}
\Hat{\Gamma}{^\gamma{}_{\mu\nu}} \equiv e{^\gamma_i}\partial_\nu e{^i_\mu}\equiv -e{^i_\mu}\partial_\nu e{^\gamma_i},
 \end{equation}
This is a simple connection that only exhibits torsion but not curvature \cite{aldrovandi1995introduction,bahamonde2023teleparallel}. This formulation implies that the Weitzenb\"ock covariant derivative of the tetrad field vanishes in the same way
 \begin{equation}\label{2}
    \nabla_\nu e^i{}_{\mu}\equiv \partial_\nu e^i{}_{\mu}- \hat{\Gamma}{^\gamma{}_{\mu\nu}}e^i{}_{\gamma}=0,
\end{equation}   
This is the condition known as absolute parallelism. Additionally, torsion related to the Weitzenb\"ock connection
is
\begin{equation}\label{3}
  T{^\gamma{}_{\mu\nu}}\equiv \Hat{\Gamma}{^\gamma{}_{\nu\mu}}-\Hat{\Gamma}{^\gamma{}_{\mu\nu}} \equiv e{^\gamma_i}(\partial_\mu e{^i_\nu}-\partial_\nu e{^i_\mu}).
\end{equation}    

This tensor encompasses all the information about the
gravitational field.
The teleparallel
equivalent of general relativity (TEGR) Lagrangian is built with the torsion equation \eqref{3}, and its dynamical equations for the vierbein imply  Einstein's equations for the metric. The
torsion scalar $(T)$ is given as
\begin{equation}\label{4}
T\equiv T{^\gamma{}_{\mu\nu}}S{_\gamma{}^{\mu\nu}},
\end{equation}
where,
\begin{equation}\label{5}
    S{_\gamma{}^{\mu\nu}}=\frac{1}{2}(K{^{\mu\nu}{}_\gamma}+\delta{^\mu_\gamma}T{^{\beta\nu}{}_\beta}-\delta{^\nu_\gamma}T{^{\beta\mu}{}_\beta}),
\end{equation}

The difference between the Levi-Civita and Weitzenb\"ock
connections is known as the contorsion tensor $K{^{\mu\nu}{}_\gamma}$ and it is defined as

\begin{equation}\label{6}
    K{^{\mu\nu}{}_\gamma}=-\frac{1}{2}(T{^{\mu\nu}{}_\gamma}-T{^{\nu\mu}{}_\gamma}-T{_\gamma{}^{\mu\nu}}).
\end{equation}

The action in $ f(T)$ gravity \cite{rodrigues2012anisotropic,amir2015kantowski,bhoyar2017stability} is defined as
\begin{equation}\label{7}
    S=\int[f(T)+L_{M}]d^4x\sqrt{-g},
\end{equation}
where $\ f(T)$ denotes an
algebraic function of the torsion scalar $T$, $ L_{M}$ represents the matter Lagrangian, and $\sqrt{-g}=det[e{^i_\mu}]=e$. The equation of motion is obtained by the functional variation of action equation $\eqref{5}$ concerning tetrads as \cite{bhoyar2017stability,amir2015kantowski,samanta2013kantowski,rodrigues2012anisotropic}
\begin{equation}\label{8}
S{_\mu{}^{\nu\rho}}(\partial_ \rho T)  f_{T T} +[e^{-1}e{^i_\mu}\partial_\rho(ee{^\gamma_i}S{_\gamma{}^{\nu\rho}}) +T{^\gamma{}_{\lambda\mu}}  S{_\gamma{}^{\nu\lambda}}]f_T  + \frac{1}{4}\delta{^\nu_\mu}f=4\pi T{^\nu_\mu}.
\end{equation}

The energy-momentum tensor $ T{^\nu_\mu}$ is considered as bulk viscous fluid coupled with a one-dimensional cosmic string. $f_T$ and $f_{TT}$ represent the first and second order derivatives of $f(T)$ with respect to the torsion scalar $ T$.

The source is a bulk viscous fluid coupled with a one-dimensional cosmic string given by \cite{xing2005bianchi,mishra2019bulk,tripathy2010anisotropic} as
\begin{equation}\label{9}
    T{^\nu_\mu}=(\rho +\overline{p})u_\mu u^\nu-\overline{p} g{^\nu_\mu}-\lambda x_\mu x^\nu,
\end{equation}
\begin{equation}\label{10}
    \overline{p}=p-3\xi H,
\end{equation}

where $\rho=\rho_p+\lambda$ is the total energy density with particles attached to them,  $\rho_p$ is the particle energy density, $\lambda$ is the string energy density, $p-3\xi H$ is the bulk viscous pressure, $\xi(t)$ is the coefficient of bulk viscosity, $H$ is Hubble parameter, $ x^\nu$ is a unit space-like vector for cloud string and $ u^\nu$ denotes the four velocities satisfying the conditions $u^\nu u_\nu=1=-x^\nu x_\nu$ and $u_\nu x^\nu=0$.

In the comoving coordinate system, we have
\begin{equation}\label{11}
    u^\nu =(0,0,0,1),\quad x^\nu = (A^{-1},0,0,0).
\end{equation}

\section{Kantowski-Sachs universe}\label{III}
The Kantowski-Sachs geometry can be viewed as the anisotropic extension of the closed Friedmann-Lema\^itre- Robertson-Walker (FLRW) geometry. The metric of this geometry is dependent on two crucial scale factors in the spacelike hypersurface. By employing Misner-like variables \cite{ryan2015homogeneous,misner1968isotropy}, one scale factor represents the radius of the spacelike hypersurface, while the second scale factor characterizes the anisotropy. When the anisotropy remains constant, the Kantowski-Sachs spacetime exhibits a six-dimensional Lie algebra as killing vector fields, and it reaches the limit of the closed FLRW geometry \cite{weber1984kantowski}. Kantowski-Sachs space is closely linked to  Bianchi spacetime \cite{ryan2015homogeneous}. The Kantowski-Sachs geometry arises from a Lie contraction in the  LRS Bianchi type-IX.

Various studies have been conducted on Kantowski-Sachs geometries due to their importance \cite{adamek2010anisotropic,collins1977global,latta2016kantowski,leon2011dynamics}. The effects of the cosmological constant on GR have been studied \cite{weber1985kantowski,gron1987dust}, while the case of a perfect fluid obeying the barotropic equation of state was examined by C.B. Collins et al. \cite{collins1977global}. The presence of the cosmological constant leads the Kantowski-Sachs universe to become a de Sitter universe in the future \cite{weber1985kantowski}, making it suitable for describing the preinflationary era \cite{gron1987dust}. In \cite{collins1977global}, it was discovered that models with a fluid source exhibit a past asymptotic behavior resulting in a Big-Bang singularity and a future attractor leading to a large crunch. The effects of the electromagnetic field were explored in \cite{dhurandhar1980electromagnetic,banerjee1999dilaton,garcia2005singularity}. The significance of the Kantowski-Sachs geometry can be observed in homogeneous models. For silent universes, this importance is clear \cite{bruni1994dynamics}. In the case of Szekeres spacetimes, \cite{szekeres1975class} the field equations of  Kantowski-Sachs restrict the dynamical variables of the field equation system this means that there are silent in homogeneous and anisotropic universes that reduce to the anisotropic Kantowski-Sachs geometry when homogenized \cite{Bonnor_Tomimura_1976}. Ghosh et al. \cite{ghosh2024lorentzian} revisited the anisotropic Kantowski-Sachs model in light of a Lorentzian path integral formalism. A. Paliathanasis \cite{paliathanasis2024phase}  performed a detailed analysis of the dynamics by investigating the stationary points and reconstructing the Kantowski-Sachs cosmological model in Weyl integrable spacetime with an ideal gas. G.G. Luciano investigated noninteracting and interacting dark energy scenarios for the Kantowski-Sachs universe \cite{luciano2023saez}.
The anisotropic and homogeneous Kantowski-Sachs model, derived from a stiff fluid and cosmological constant was investigated by A. Singh for several potential universe scenarios \cite{singh2023qualitative}.

In this case,
we consider the spatially homogeneous and anisotropic Kantowski-Sachs spacetime \cite{rodrigues2012anisotropic,amir2015kantowski,samanta2013kantowski} of the 
form
\begin{equation}\label{12}
ds^2=dt^2-A^2(t)dr^2-B^2(t)(d\theta^2+\sin^2{\theta}d\phi^2),
\end{equation}

where the metric potentials $ A$ and $\ B$ are functions of cosmic time $ t$ only. Now the set of diagonal tetrads is related to the metric equation $\eqref{12}$ and is given as
\begin{equation}\label{13}
    [e{^i_\mu}]=diag[1,A,B,B\sin{\theta}].
\end{equation}

The determinant of equation \eqref{12} is
\begin{equation}\label{14}
    e=AB^2\sin{\theta}.
\end{equation}

The torsion scalar $T $ is derived from equation \eqref{4} as 
\begin{equation}\label{15}
    T=-2\biggl(\frac{2\Dot{A}\Dot{B}}{AB}+\frac{\Dot{B}^2}{B^2}\biggr).
\end{equation}

The field equations for the Kantowski-Sachs model equation\eqref{12} from equations \eqref{7}-\eqref{11} in $ f(T)$ gravity is given as follows:
\begin{equation}\label{16}
\begin{split}
  f++4\biggl(\frac{\Dot{A}\Dot{B}}{AB}+\frac{\Ddot{B}}{B}+\frac{ \Dot{B}^2}{B^2} \biggr )f_T +4\biggl(\frac{\Dot{B}}{B}\biggr) \Dot{T} f_{TT} = -16\pi(\Bar{p}-\lambda),
\end{split}
\end{equation}

\begin{equation}\label{17}
\begin{split}
  f +2\biggl(3\frac{\Dot{A}\Dot{B}}{AB}+ \frac{\Ddot{A}}{A} + \frac{\Ddot{B}}{B}+  \frac{ \Dot{B}^2}{B^2}  \biggr )f_T   +2\biggl(\frac{\Dot{A}}{A}+\frac{\Dot{B}}{B} \biggr) \Dot{T} f_{TT}=-16\pi \Bar{p},
\end{split}
\end{equation}
    
\begin{equation}\label{18}
     f+4\biggl(2\frac{\Dot{A}\Dot{B}}{AB}+\frac{\Dot{B}^2}{B^2} \biggr)f_T=16\pi\rho,
\end{equation} 

 where the overhead dot $(.)$ represents the derivative concerning cosmic time $t$. $ f_T$ and $f_{TT}$ are the first and second order derivatives of $f(T)$ with respect to $ t$, respectively. We have three nonlinear differential equations and six unknowns namely $ f$, $A$, $ B$, $\Bar{p} $, $ \rho$, and $\lambda$. The volume $V$ and average scale factor $a(t)$ are defined as
 
 \begin{equation}\label{19}
     V=[a(t)]^3=AB^2.
 \end{equation}
 The anisotropic parameter $(A_m)$ is given by
 \begin{equation}\label{20}
     A_m=\frac{1}{3}\sum_{i=1}^{3}\biggl(\frac{H_i-H}{H}\biggr)^2,
 \end{equation}
where $H_1=\frac{\Dot{A}}{A}$, and $H_2=\frac{\Dot{B}}{B}=H_3$  are directional Hubble parameters, and the mean Hubble parameter is defined as

\begin{equation}\label{21}
    H=\frac{\Dot{a}}{a}=\frac{1}{3}\biggl(\frac{\Dot{A}}{A}+2\frac{\Dot{B}}{B}\biggr).
\end{equation}

The expansion scalar $(\theta)$ and shear scalar $(\sigma)$ are defined as
\begin{equation}\label{22}
    \theta=u^i_{;i}=\frac{\Dot{A}}{A}+2\frac{\Dot{B}}{B},
\end{equation}
\begin{equation}\label{23}
    \sigma^2=\frac{3}{2}H^2A_m.
\end{equation}

The deceleration parameter $q$ is given as
\begin{equation}\label{24}
    q=-\frac{\Ddot{a}a}{\Dot{a}^2}=-1+\frac{d}{dt}\biggl(\frac{1}{H}\biggr).
\end{equation}

\section{Solution of field equations}\label{IV}

\begin{figure}[H]

\begin{subfigure}{0.5\textwidth}
\includegraphics[width=0.9\linewidth,
height=6cm]{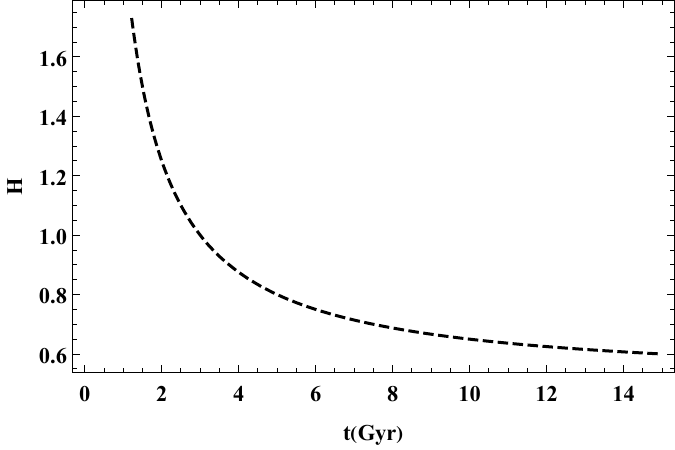} 
\caption{}
\label{I a}
\end{subfigure}
\begin{subfigure}{0.5\textwidth}
\includegraphics[width=0.9\linewidth, height=6cm]{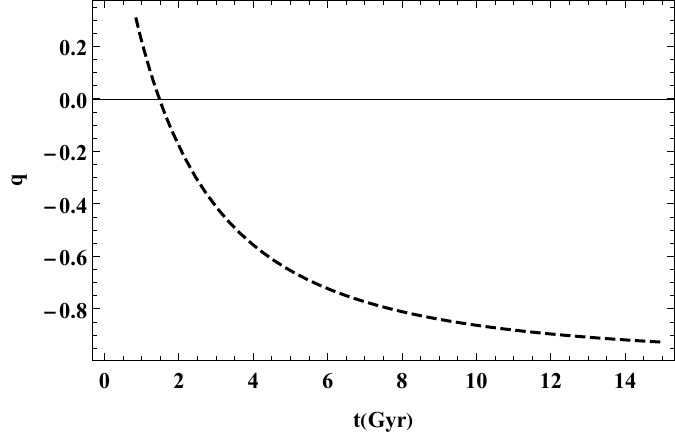}
\caption {}
\label{I b}
\end{subfigure}
\caption{These figures show the behavior of  Hubble parameter $H$ and deceleration parameter $q$ at $n=1.5$, $b=0.25$ and $t_0=0.5$.}
\label{fig:I}
\end{figure}

To solve nonlinear differential field equations \eqref{16}-\eqref{18}, the following physically conceivable scenarios are considered \cite{akarsu2014cosmology}.
\begin{enumerate}[I.]
    \item The hybrid expansion law (HEL), which is proposed as 

\begin{equation}\label{25}
    a(t)=a_0\biggl(\frac{t}{t_0}\biggr)^n e^{b(\frac{t}{t_0}-1)},
\end{equation}

 where $a_0$ and $t_0$  denote the scale factor and age of the today's universe, respectively, and $n >0 $ and $b> 0$ are the free parameters. As previously noted, the HEL has been utilized to solve the field equation. Only epoch-based universe evolution can be explained by the power-law and exponential-law cosmologies due to the consistency of the deceleration parameter. The universe does not seem to be shifting from a state of deceleration to acceleration in these cosmologies. Kumar \cite{kumar2013anisotropic} and Akarsu et al. \cite{akarsu2014cosmology} take into consideration the following form of the universe's scale factor to explain such a transition.
 The HEL, also known as the generalized form of the scale factor, gives rise to a power-law cosmology for $b=0 $  and an exponential law for $n=0$, while the combination of   $n >0 $ and $b> 0$ results in a new cosmology arising from the HEL. By choosing this scale factor, we obtain a time-dependent deceleration parameter (eqn. \eqref{35}  and fig. \eqref{I a}) that represents inflation and a radiation/matter-dominated era before the dark energy era, followed by a transition from deceleration to acceleration.

\item 

\begin{equation}\label{26}
    A=B^k, 
\end{equation}

where $k$ is constant, as the observed velocity of the redshift relation beyond the galactic sources suggests that the Hubble expansion of the universe is isotropic within a range of 30$\%$. In a more precise manner, studies on redshift have established a limit of  $\sigma/H\leq 0.3$   on the ratio of shear $\sigma$ to the Hubble constant $H$ in the vicinity of our galaxy. Collin et al. \cite{collins1980exact} and Y. Aditya et al. \cite{prasanthi2020anisotropic} highlighted that for a spatially homogeneous metric, the normal congruence to homogeneous expansion fulfils the condition that $\sigma/\theta$ remains constant. In other words, the expansion scalar is directly proportional to the shear scalar, which provides the relationship between the metric potentials \eqref{26}.

    \item As noted  by \cite{reddy2013anisotropic} and \cite{naidu2013bianchi} the combined 
effect of the bulk viscous pressure and proper pressure 
is expressed as 
\begin{equation}\label{27}
\Bar{p}=p-3\xi H = \varepsilon\rho,
\end{equation}

where
\begin{equation}\label{28}
    \varepsilon=\varepsilon_0 - \gamma \quad (0\leq \varepsilon_0 \leq 1), \quad p=\varepsilon_0\rho,
\end{equation}
 $\varepsilon_0$, and $\gamma$ are constant.
\end{enumerate}
 The values of the metric potentials $ A$ and $\ B$ are obtained from equations \eqref{19}, \eqref{25}, and \eqref{26} as
   \begin{equation}\label{29}
    A=\biggl[a_0\biggl(\frac{t}{t_0}\biggr)^n e^{b(\frac{t}{t_0}-1)}\biggr]^\frac{3k}{k+2},
\end{equation}
\begin{equation}\label{30}
    B=\biggl[a_0\biggl(\frac{t}{t_0}\biggr)^n e^{b(\frac{t}{t_0}-1)}\biggr]^\frac{3}{k+2}.
\end{equation}

Now, the metric equation \eqref{12} is written with the help of equations \eqref{29} and \eqref{30} as
\begin{equation}\label{31}
\begin{split}
   ds^2=dt^2-\biggl[a_0\biggl(\frac{t}{t_0}\biggr)^n e^{b(\frac{t}{t_0}-1)}\biggr]^\frac{3k}{k+2}dr^2 - \biggl[a_0\biggl(\frac{t}{t_0}\biggr)^n e^{b(\frac{t}{t_0}-1)}\biggr]^\frac{3}{k+2}\biggl(d\theta^2-\sin^2{\theta}d\phi^2\biggr). 
   \end{split}
\end{equation}

The volume $V$ of the model is given by equations \eqref{19} and \eqref{25} as

\begin{equation}\label{32}
    V= \biggl[a_0\biggl(\frac{t}{t_0}\biggr)^n e^{b(\frac{t}{t_0}-1)}\biggr]^3.
\end{equation}

The coefficient of the bulk viscosity parameter $\xi$ is obtained from equations \eqref{27} and \eqref{28}  as
\begin{equation}\label{33}
    \xi=\frac{\rho\gamma}{3H}.
\end{equation}
The mean Hubble parameter $H$ and deceleration $q$ are given by equations \eqref{21}, \eqref{24}, and \eqref{25} as
\begin{equation}\label{34}
    H=\frac{n}{t}+\frac{b}{t_0},
\end{equation}

\begin{equation}\label{35}
    q=-\frac{b^2t^2+2bntt_0+n(n-1)t^2_0}{(bt+nt_0)^2}.
\end{equation}


The graphical depictions in Fig. \eqref{I a} showcase the behavior of  Hubble parameter $H$. The Hubble parameter exhibits a discernible trend, and its value decreases as cosmic time $t$ progresses. The consistently positive values of $H$ throughout the cosmic evolution unequivocally signify the ongoing expansion of the universe. The Hubble parameter gradually converges toward the smallest positive value.

The deceleration parameter $q$. displayed in Fig. \eqref{I b}, changes over time, showing a transition from the early decelerating phase to the late accelerating phase.  This pattern nicely matches what we observe in Type-I supernovae, as  $q<0$  for   $\frac{t_0(\sqrt{n}-n)}{b}< t$. Hence, the universe has entered an accelerating phase with a deceleration parameter ranging between $ - 1 \leq q \leq 0$. 
These results agree well with \cite{r31,q2mishra2019bulk,q4yadav2012bianchi,q5bharali2023dynamics}.

The  power law model  of $f(T)$ gravity is given  as 
\begin{equation}\label{36}
    f(T)=\alpha T +\beta T^m , \quad m\geq2,
\end{equation}
\quad where $\alpha$ and $\beta$ are model parameters.
Since the chosen form of $f(T)$ gravity is nonlinear, it allows for a more intricate and complex description of the gravitational dynamics compared to linear or simpler forms. This nonlinearity can capture more intricate gravitational effects and behaviors. We consider three models based on the following model parameter options.
\begin{equation*}
        i)\quad \alpha \neq 0, \beta=0,
        \quad
         ii)\quad \alpha=0, \beta\neq 0,
         \quad   iii)\quad \alpha \neq0, \beta \neq 0. 
\end{equation*}

By varying these parameters, we can explore various scenarios and study how different values of model parameters impact cosmological evolution. This approach allows for exploring how torsion scalar terms with different powers contribute to the gravitational field equations and influence the universe's dynamics \cite{gadbail2021power,bhoyar2017stability,paliathanasis2016cosmological}.

\subsection*{Model I: $\alpha \neq 0$ and $\beta=0$}
In the first case, we constrained the values of the model parameters to $\alpha=1$ and $\beta=0$. With the help of equations \eqref{29} and \eqref{30}, the field equations  \eqref{16}-\eqref{18} are obtained as

\begin{equation}\label{37}
     f+4\biggl[\frac{B\Ddot{B}+(k+1)\Dot{B}^2}{B^2}\biggr]   =-16\pi( \Bar{p}-\lambda),
\end{equation}
\begin{equation}\label{38}
     f+2(1+k)\biggl[\frac{B\Ddot{B}+(k+1)\Dot{B}^2}{B^2}\biggr]    =-16\pi \Bar{p},
\end{equation}
\begin{equation}\label{39}
  f+4\biggl[(2k+1)\frac{\Dot{B}^2}{B^2}\biggr]    =16\pi\rho.   
\end{equation}

\subsection*{Model II: $\alpha=0$ and $\beta\neq 0$}
In this case, we constrained the model parameters to $\alpha=0$ and $\beta=1$. The field equations are obtained from Eqs. \eqref{29}and \eqref{30} then Eqs. \eqref{16}-\eqref{18} are rewritten as
\begin{equation}\label{40}
   f+8T\biggl[\frac{B\Ddot{B}+(k+1)\Dot{B}^2}{B^2}\biggr]  +8\Dot{T}\frac{\Dot{B}}{B}=-16\pi( \Bar{p}-\lambda),  
\end{equation}
\begin{equation}\label{41}
    f+4(k+1)T\biggl[\frac{B\Ddot{B}+(k+1)\Dot{B}^2}{B^2}\biggr]  +4\Dot{T}\frac{\Dot{B}}{B}(k+1)  =-16\pi \Bar{p},
\end{equation}
\begin{equation}\label{42}
     f+8\Dot{T}\biggl[(2k+1)\frac{\Dot{B}^2}{B^2}\biggr]   =16\pi\rho.
\end{equation}

\subsection*{Model III: $\alpha\neq 0$ and $\beta\neq 0$}
In this case, we consider the values of the model parameters $\alpha=1$ and $\beta=-1$. Rewriting the field Eqs. \eqref{16}-\eqref{18} with the help of Eqs. \eqref{29} and \eqref{30}, we obtain the new field equations as

\begin{equation}\label{43}
    f+4(1-2T)\biggl[\frac{B\Ddot{B}+(k+1)\Dot{B}^2}{B^2}\biggr]  +8\Dot{T}\frac{\Dot{B}}{B}=-16\pi( \Bar{p}-\lambda),  
\end{equation}
\begin{equation}\label{44}
    f+2(1-2T)(1+k)\biggl[\frac{B\Ddot{B}+(k+1)\Dot{B}^2}{B^2}\biggr] +4\Dot{T}\frac{\Dot{B}}{B}(k+1)=-16\pi \Bar{p},
\end{equation}
\begin{equation}\label{45}
     f+4(1-2T)\biggl[(2k+1)\frac{\Dot{B}^2}{B^2}\biggr] =16\pi\rho.
\end{equation}

To obtain the value of  $\rho$ for the models I to III, we use the field Eqs. \eqref{39}, \eqref{42} and \eqref{45}. The total energy density for  each model is calculated as

\begin{equation}\label{46}
    \rho=\frac{1}{16\pi}\Biggl\{f+4\biggl[(2k+1)\frac{\Dot{B}^2}{B^2}\biggr]\Biggr\},
\end{equation}

\begin{equation}\label{47}
    \rho=\frac{1}{16\pi}\Biggl\{f+8\Dot{T}\biggl[(2k+1)\frac{\Dot{B}^2}{B^2}\biggr]\Biggr\},
\end{equation}

\begin{equation}\label{48}
    \rho=\frac{1}{16\pi}\Bigg\{f+4(1-2T)\biggl[\frac{B\Ddot{B}+(k+1)\Dot{B}^2}{B^2}\biggr]  +8\Dot{T}\frac{\Dot{B}}{B}\Bigg\}.
\end{equation}

\begin{figure}[htbp]
\centering
\begin{subfigure}[b]{0.33\textwidth}
\centering
    \includegraphics[width=\textwidth]{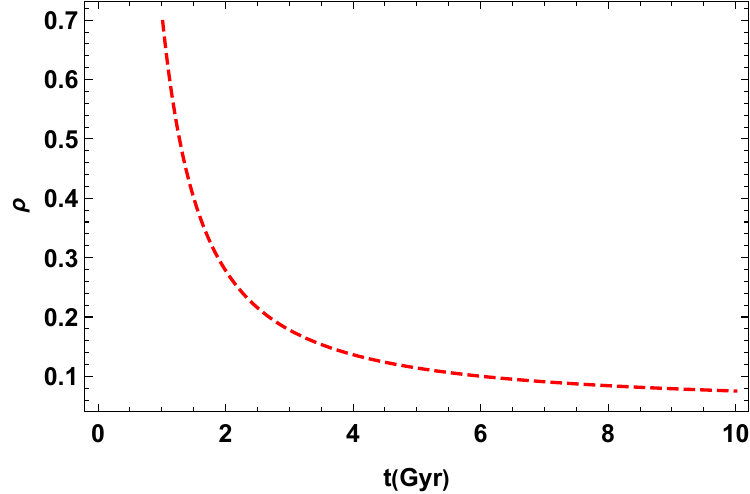}
\caption{Model-I}
\label{II a}
\end{subfigure}
\hfill
\begin{subfigure}[b]{0.33\textwidth}
        \centering
        \includegraphics[width=\textwidth]{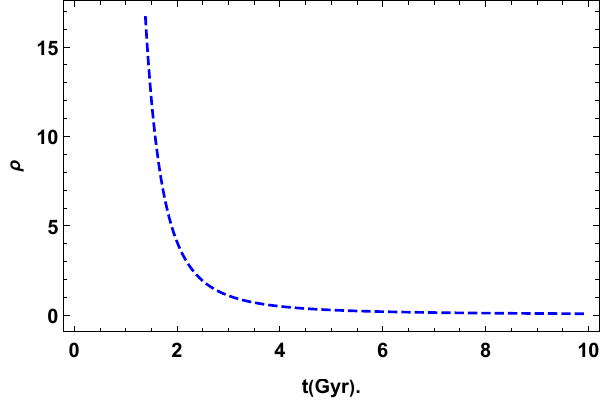}
\caption{Model-II}
\label{II b}
    \end{subfigure}
    \hfill
\begin{subfigure}[b]{0.33\textwidth}    \centering
\includegraphics[width=\textwidth]{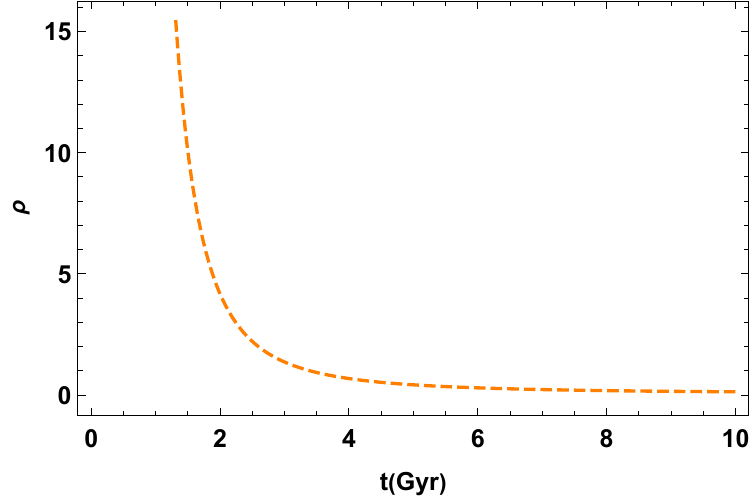}
\caption{Model-III}
\label{II c}
\end{subfigure}

\caption{These figures show the behavior of the total energy density
$\rho$ at $n=1.5$, $b=0.25$, $t_{0}=0.5$ and $k=0.8$ for models-I, II, and III respectively.}
    \label{fig:II}
\end{figure}

The proper energy density $\rho$ exhibits a swift decline from positive values to the same constant, as shown in Figs. \eqref{II a}, \eqref{II b}, and \eqref{II c} for models I, II, and III, respectively. The substantial initial values of $\rho$  signify the domination of densities during the early stages, which decreases significantly as time progresses. Importantly, $\rho$ maintains positive values throughout the entire evolution of the universe. This trend underscores the expansive nature of the universe. According to \cite{bhoyar2017stability}, $\rho \rightarrow 0$, for $ t \rightarrow \infty$  the universe approaches a flat universe at a late time. As a result, each model is in good agreement with recent observations, and these respondents strongly agree with \cite{r30,r2moreira2023string,r3sahoo2016bianchi}.

The bulk viscous pressure $\Bar{p}$ of models I to III  are obtained from field Eqs. \eqref{38}, \eqref{41}, and \eqref{44} as

\begin{equation}
    \Bar{p}=-\frac{1}{16\pi}\Bigg\{f+2(1+k)\biggl[\frac{B\Ddot{B}+(k+1)\Dot{B}^2}{B^2}\biggr]\Bigg\},
\end{equation}

\begin{equation}
    \Bar{p}=-\frac{1}{16\pi}\Bigg\{f+4(k+1)T\biggl[\frac{B\Ddot{B}+(k+1)\Dot{B}^2}{B^2}\biggr] +4\Dot{T}\frac{\Dot{B}}{B}(k+1)\Bigg\},
\end{equation}

\begin{equation}\label{51}
    \Bar{p}=- \frac{1}{16\pi}\Bigg\{ f+2(1-2T)(1+k)\biggl[\frac{B\Ddot{B}+(k+1)\Dot{B}^2}{B^2}\biggr] +4\Dot{T}\frac{\Dot{B}}{B}(k+1)\Bigg\}.
\end{equation}

\begin{figure}[H]
\centering
\begin{subfigure}[b]{0.33\textwidth}
\centering
    \includegraphics[width=\textwidth]{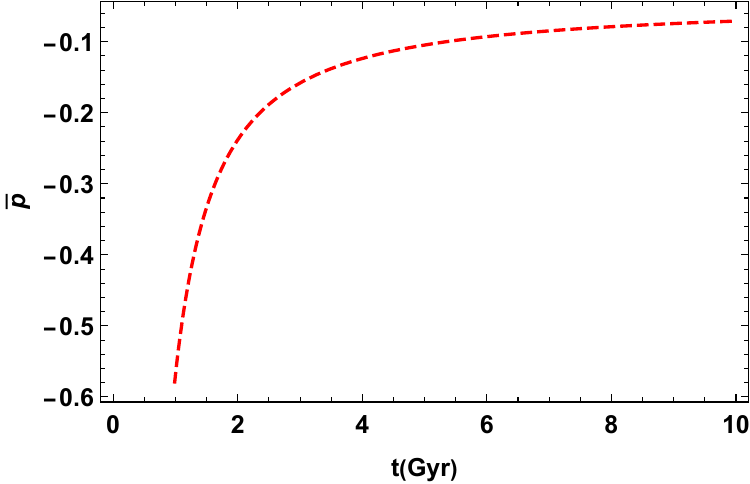}
\caption{Model-I}
\label{III a}
\end{subfigure}
\hfill
\begin{subfigure}[b]{0.33\textwidth}
        \centering
        \includegraphics[width=\textwidth]{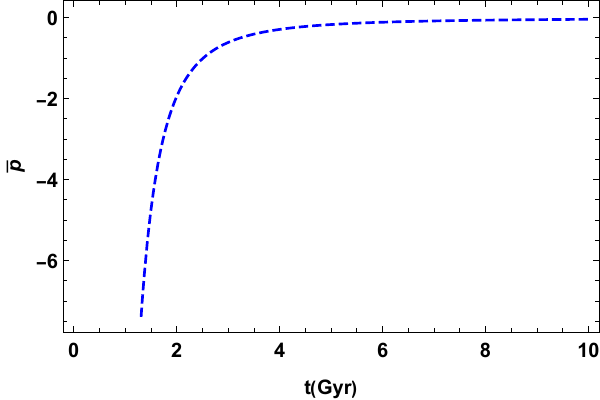}
\caption{Model-II}
\label{III b}
    \end{subfigure}
    \hfill
\begin{subfigure}[b]{0.33\textwidth}    \centering
\includegraphics[width=\textwidth]{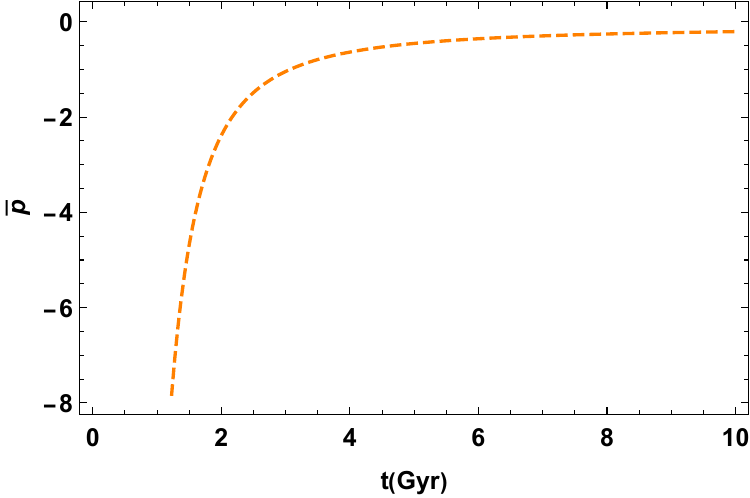}
\caption{Model-III}
\label{III c}
\end{subfigure}

\caption{These figures show the behavior of the effective pressure
$\Bar{p}$ at $n=1.5$, $b=0.25$, $t_{0}=0.5$ and $k=0.8$ for models- I, II, and III, respectively.}
    \label{fig:III}
\end{figure}

The examination of Figs \eqref{III a}, \eqref{III b} and \eqref{III c} reveals a noteworthy trend for models I, II, and III respectively in the behavior of the bulk viscous ${\Bar{p}}$ pressure across the cosmic evolution. The parameter ${\Bar{p}}$ consistently exhibits a negative trajectory throughout this evolution, beginning with high negative values later on the models II and III show $\Bar{p}\rightarrow 0$, while model I  approach a small negative value. This characteristic behavior implies the accelerated expansion of the universe. The persistent negativity in bulk viscous pressure signifies a sustained repulsive effect, contributing to the observed cosmic acceleration; these arguments hold similar results to \cite{r30,r2moreira2023string,r3sahoo2016bianchi,amirhashchi2011magnetized,amirhashchi2013string}.

The effective equation of state ($\omega_{eff}=\rho/\Bar{p} $) for each model is calculated from Eqs. \eqref{46} to \eqref{51} as

\begin{equation}
    \omega_{eff}=- \frac{B^2 f +2(1+k)[B\Ddot{B}+(k+1)\Dot{B}^2]}{B^2 f +4(2k+1)\Dot{B}^2},
\end{equation}

\begin{equation}
    \omega_{eff}=-\frac{B^2f+4(1+k)[B\Ddot{B}+(k+1)\Dot{B}^2] T+ 4B(k+1)\Dot{B}\Dot{T}}{B^2f+8(2k+1)\Dot{B}^2\Dot{T}},
\end{equation}
\begin{equation}
    \omega_{eff}=-\frac{B^2f+2(1-2T)(1+k)[B\Ddot{B}+(k+1)\Dot{B}^2]+4(k+1)B\Dot{B}\Dot{T}}{B^2f+4(1-2T)[B\Ddot{B}+(k+1)\Dot{B}^2]+8B\Dot{B}\Dot{T}}.
\end{equation}

\begin{figure}[H]
\centering
\begin{subfigure}[b]{0.33\textwidth}
\centering
    \includegraphics[width=\textwidth]{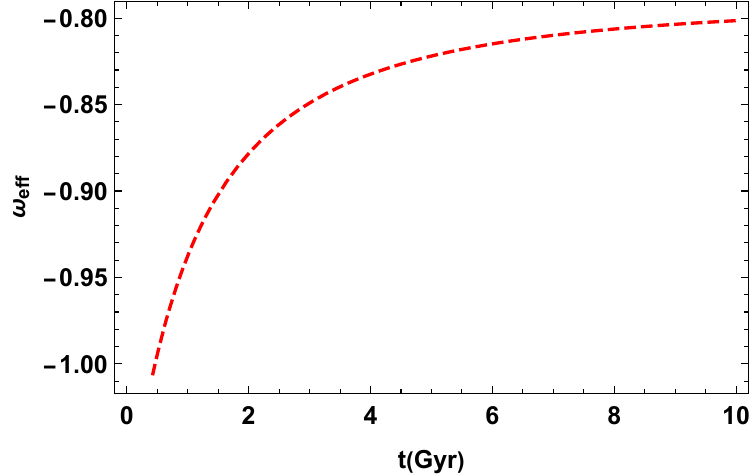}
\caption{Model-I}
\label{IV a}
\end{subfigure}
\hfill
\begin{subfigure}[b]{0.33\textwidth}
        \centering
        \includegraphics[width=\textwidth]{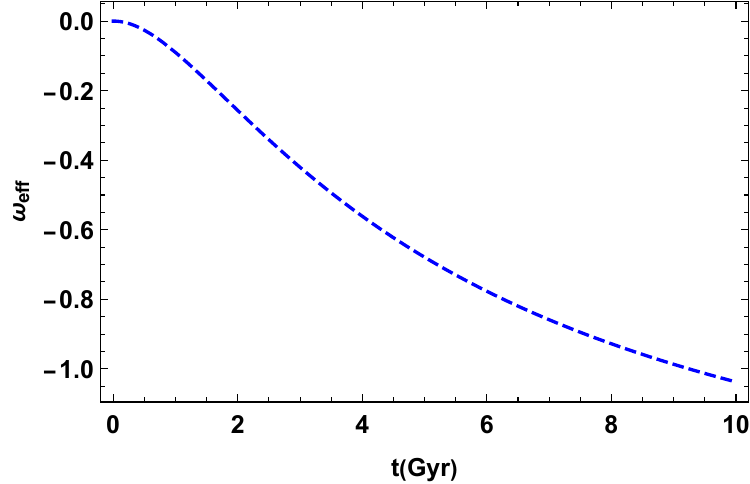}
\caption{Model-II}
\label{IV b}
    \end{subfigure}
    \hfill
\begin{subfigure}[b]{0.33\textwidth}    \centering
\includegraphics[width=\textwidth]{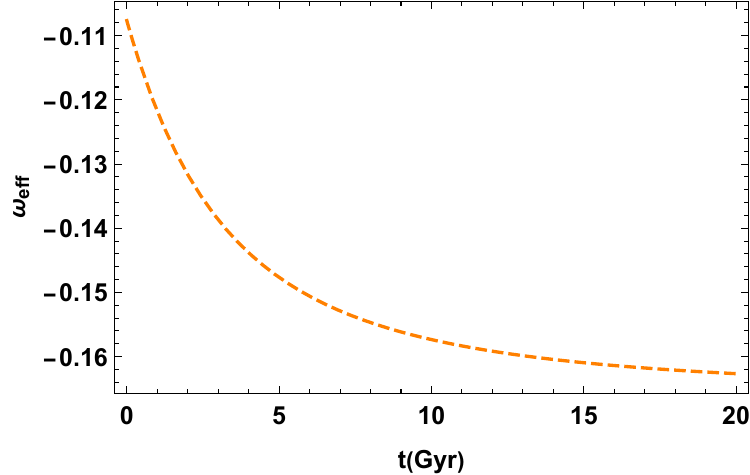}
\caption{Model-III}
\label{IV c}
\end{subfigure}

\caption{These figures show the behavior of the effective equation of state
$\omega_{eff}$ at $n=1.5$, $b=0.25$, $t_{0}=0.5$ and $k=0.8$ for models-I, II, and III, respectively.}
    \label{fig:IV}
\end{figure}

One of the largest endeavours in observational cosmology is the measurement of the EoS parameter for DE. The traditional description of the DE model is provided by the effective EoS parameter $\omega_{eff}=\Bar{p}/\rho $, which is not necessarily constant. Figure \eqref{fig:IV}  portrays the behavior of the  EoS parameter along with cosmic time using the hybrid expansion law for some fixed values of $n$, $b$, and $t_{0}$.  

According to, observational data  $\omega_{eff}$ should remain on the negative axis and less than $-1/3$ for model I. To remain on the negative axis, we choose the negative value of the model parameter $\beta=-1$. We observed that (Fig. \eqref{IV a} $),  \omega_{eff}$ for model I decreases from a higher negative value to a smaller negative value. However, for models II and III (Fig. \eqref{IV b},\eqref{IV c}), it becomes reversible, i.e., $\omega_{eff}$ increases from a smaller negative to a higher negative value. For models I and II,  $\omega_{eff}\leq -1$. In other words, models I and II remain in the quintessence region, but model II approaches $-1$ as time $t$ progresses; hence, it is possible that model II behaves like a standard $\Lambda$CDM. This finding is in good agreement with \cite{r3sahoo2016bianchi,santhi2022bulk,rodrigues2015bianchi,mishra2019bulk}.

The string energy density $\lambda$ for each model is obtained from the field equations \eqref{37}, \eqref{38}, \eqref{40}, \eqref{41}, \eqref{43}, and \eqref{44} as

\begin{equation}
    \lambda=\frac{1}{16\pi}\Biggl\{f+4\biggl[\frac{B\Ddot{B}+(k+1)\Dot{B}^2}{B^2}\biggr]\Biggr\}+\Bar{p},
\end{equation}

\begin{equation}
    \lambda=\frac{1}{16\pi}\Bigg\{f+8T\biggl[\frac{B\Ddot{B}+(k+1)\Dot{B}^2}{B^2}\biggr]  +8\Dot{T}\frac{\Dot{B}}{B}\Bigg\}+\Bar{p},
\end{equation}

\begin{equation}
    \lambda=\frac{1}{16\pi}\Biggl\{ f+4(1-2T)\biggl[\frac{B\Ddot{B}+(k+1)\Dot{B}^2}{B^2}\biggr]  +8\Dot{T}\frac{\Dot{B}}{B}\Bigg\}+\Bar{p}.
\end{equation} 

\begin{figure}[H]
\centering
\begin{subfigure}[b]{0.33\textwidth}
\centering
    \includegraphics[width=\textwidth]{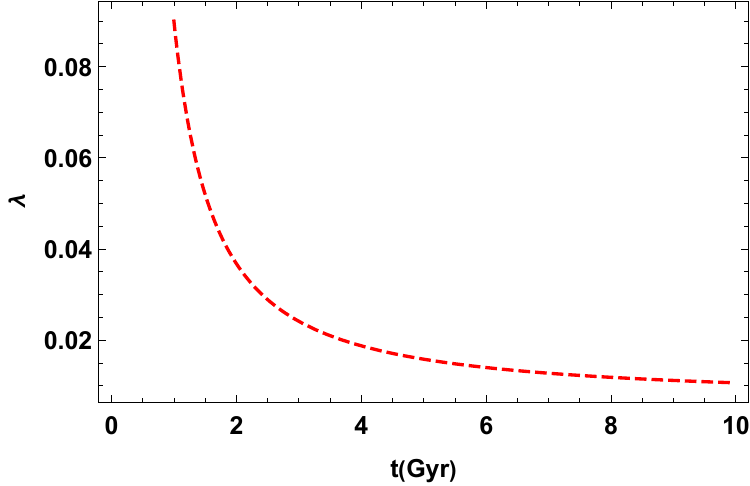}
\caption{Model-I}
\label{V a}
\end{subfigure}
\hfill
\begin{subfigure}[b]{0.33\textwidth}
        \centering
        \includegraphics[width=\textwidth]{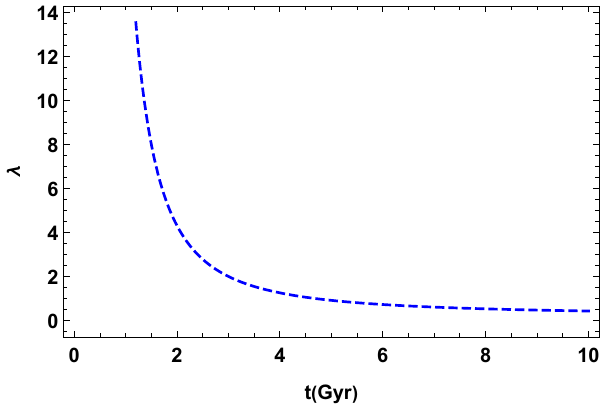}
\caption{Model-II}
\label{V b}
    \end{subfigure}
    \hfill
\begin{subfigure}[b]{0.33\textwidth}    \centering
\includegraphics[width=\textwidth]{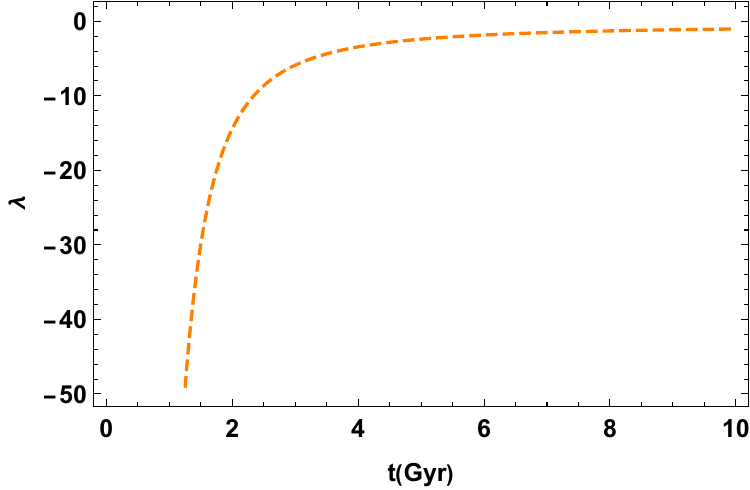}
\caption{Model-III}
\label{V c}
\end{subfigure}

\caption{These figures show the behavior of the string energy density
$\lambda$ at $n=1.5$, $b=0.25$, $t_{0}=0.5$ and $k=0.8$ for models-I, II, and III, respectively.}
    \label{fig:V}
\end{figure}

We have noted that the string energy density $\lambda$ of models I and II (Fig. \eqref{V a} and \eqref{V b}) exhibits a positive correlation with time, always maintaining a positive value until it eventually converges to zero at a subsequent point in time. Letelier \cite{letelier1983string} highlighted the potential occurrence of both positive and negative values for $\lambda$. In model III  (Fig. \eqref{V c}), $\lambda$ remains negative throughout the evolution and converges to zero for a late time. For model III where $\lambda<0$, the string phase of the universe ceases to exist, resulting in the presence of an anisotropic fluid composed of particles. Our findings show comparable outcomes to those of \cite{amirhashchi2011magnetized,amirhashchi2013string}.

The coefficient of bulk viscosity $\xi$ for the models are obtained from Eqs. \eqref{33} and \eqref{46}-\eqref{48} as
\begin{equation}
    \xi= \frac{\gamma}{48\pi H}\Biggl\{f+4\biggl[(2k+1)\frac{\Dot{B}^2}{B^2}\biggr]\Biggr\},
\end{equation}

\begin{equation}
    \xi= \frac{\gamma}{48\pi H}\Biggl\{f+8\Dot{T}\biggl[(2k+1)\frac{\Dot{B}^2}{B^2}\biggr]\Biggr\},
\end{equation}

\begin{equation}
    \xi=\frac{\gamma}{48 \pi H}\Bigg\{f+4(1-2T)\biggl[\frac{B\Ddot{B}+(k+1)\Dot{B}^2}{B^2}\biggr]  +8\Dot{T}\frac{\Dot{B}}{B}\Bigg\}.
\end{equation}

\begin{figure}[H]
\centering
\begin{subfigure}[b]{0.33\textwidth}
\centering
    \includegraphics[width=\textwidth]{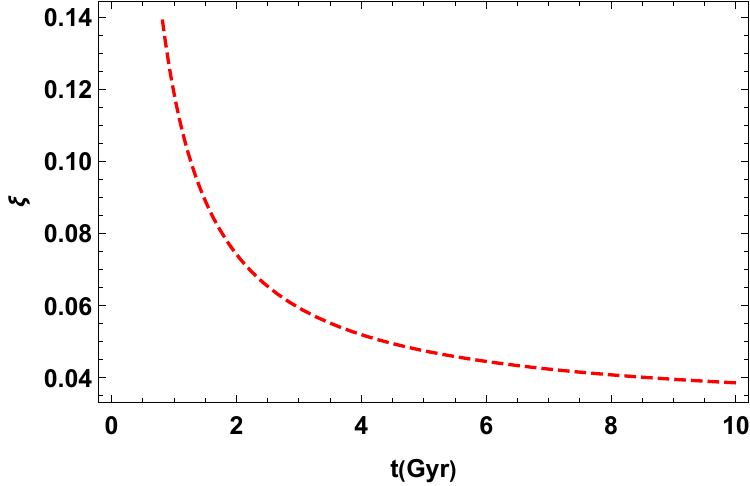}
\caption{Model I}
\label{VI a}
\end{subfigure}
\hfill
\begin{subfigure}[b]{0.33\textwidth}
        \centering
        \includegraphics[width=\textwidth]{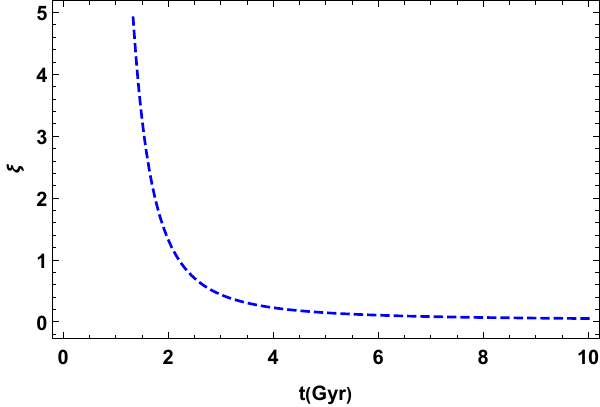}
\caption{Model II}
\label{VI b}
    \end{subfigure}
    \hfill
\begin{subfigure}[b]{0.33\textwidth}    \centering
\includegraphics[width=\textwidth]{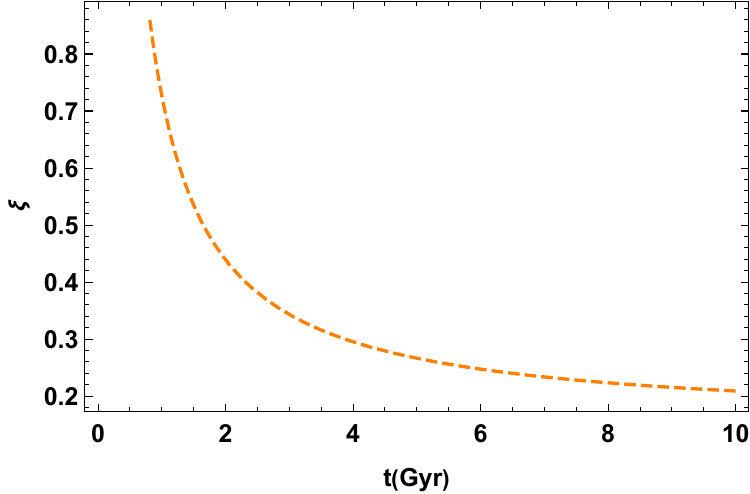}
\caption{Model III}
\label{VI c}
\end{subfigure}

\caption{The behavior of the coefficient of bulk viscosity
($\xi$) at $n=1.5$, $b=0.25$, $t_{0}=0.5$ and $k=0.8$ for models-I,II, and III, respectively.}
    \label{fig:VI}
\end{figure}

The coefficient of bulk viscosity  $\xi$ consistently decreases with time, maintaining a positive value throughout cosmic evolution for each model, as shown in Figs. \eqref{VI a}, \eqref{VI b}, and \eqref{VI c}. They ultimately lead to the emergence of inflationary models. This observed behavior aligns well with the anticipated physical characteristics of $\xi$, and these reports are in good agreement with \cite{1reddy2013anisotropic,pawar2022bulk}.

The anisotropic parameter $A_m$, expansion scalar $\theta$, and shear scalar $\sigma^2$ are given as

\begin{equation}\label{61}
    A_m=\frac{2(k-1)^2}{(k+2)^2},
\end{equation}

\begin{equation}\label{62}
    \theta=\frac{3(bt+nt_0)}{tt_0},
\end{equation}
\begin{equation}\label{63}
    \sigma^2=\frac{3(k-1)(bt+nt_0)^2}{(k+2)^2t^2t_0^2}.
\end{equation}

\begin{figure}[H]
\centering
\begin{subfigure}[b]{0.32\textwidth}
\centering
    \includegraphics[width=\textwidth]{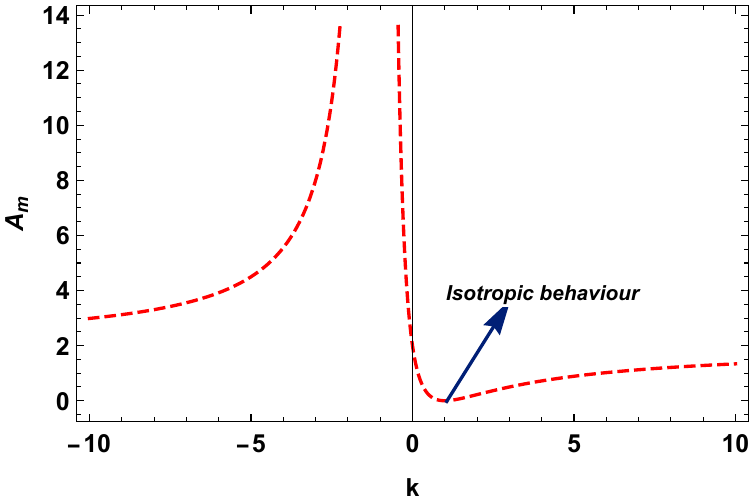}
\caption{}
\label{VII a}
\end{subfigure}
\hfill
\begin{subfigure}[b]{0.31\textwidth}
        \centering
        \includegraphics[width=\textwidth]{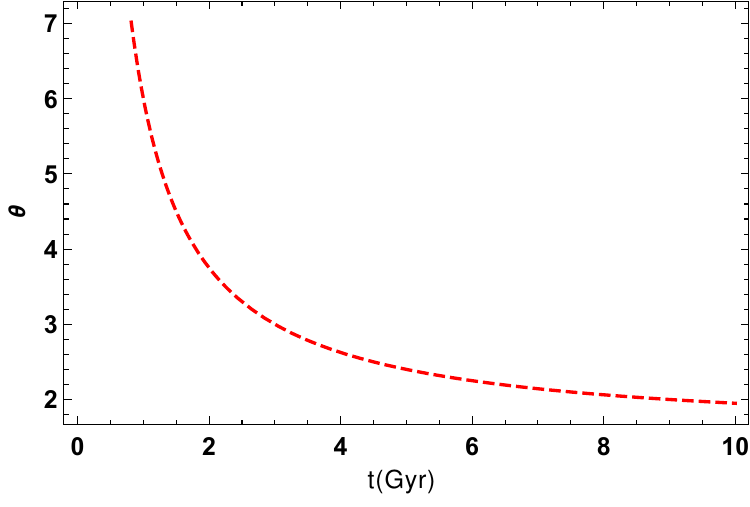}
\caption{}
\label{VII b}
    \end{subfigure}
    \hfill
\begin{subfigure}[b]{0.33\textwidth}    \centering
\includegraphics[width=\textwidth]{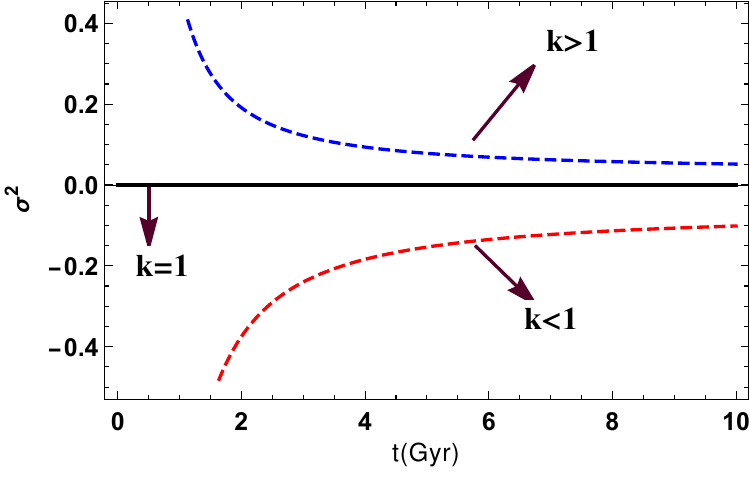}
\caption{}
\label{VII c}
\end{subfigure}

\caption{These figures show the behavior of the anisotropic parameter $A_m$, expansion scalar $\theta$, and shear scalar $\sigma^2$
 for $n=1.5$, $b=0.25$, and $t_{0}=0.5$ .}
    \label{fig:VIII}
\end{figure}

Equation \eqref{61} implies that the anisotropic parameter $A_m$ is time-independent but depends on the values of the parameter $k$, as shown in Fig. \eqref{VII a}. It is observed that the anisotropic parameter has a singularity at $k=-2$, and for $k=1$, it shows isotropic behavior. However, it becomes anisotropic when  $k>1$.

The expansion scalar $\theta$ is a positive anti-exponential time-dependent parameter that converges to zero as time $t$ progresses, as shown in Fig. \eqref{VII b}. Like the expansion scalar $\theta$, the shear scalar $\sigma^2$ is also time-dependent and similar to the anisotropic parameter $A_m$, it depends on the value of the parameter $k$, which is determined from Eqn. \eqref{63} and Fig. \eqref{VII c}. Figure \eqref{VII c} shows that for any value of ($k\neq -2$), $\sigma^2 \rightarrow 0$ as $t \rightarrow \infty$ but for $k>1$, it approaches a positive trajectory and for $k<1$, it approaches a negative trajectory.

\section{Energy Conditions}\label{V}

Numerous methods are employed to determine the universe's history and are used in many approaches to understand the evolution of the universe. Energy conditions play a crucial role in confirming hypotheses about the singularity of space-time and black holes \cite{wald2010general}, including those put forth by Roger Penrose and Stephen Hawking  \cite{ellis2014stephen}. Verifying the universe's expansion is the primary goal of these energy conditions in this context. Energy conditions can take many different forms, including dominant energy conditions (DECs), weak energy conditions (WECs), null energy conditions (NECs), and strong energy conditions (SECs) \cite{r5koussour2022bianchi}.

\begin{figure}[H]
\centering
\begin{subfigure}[b]{0.33\textwidth}
\centering
    \includegraphics[width=\textwidth]{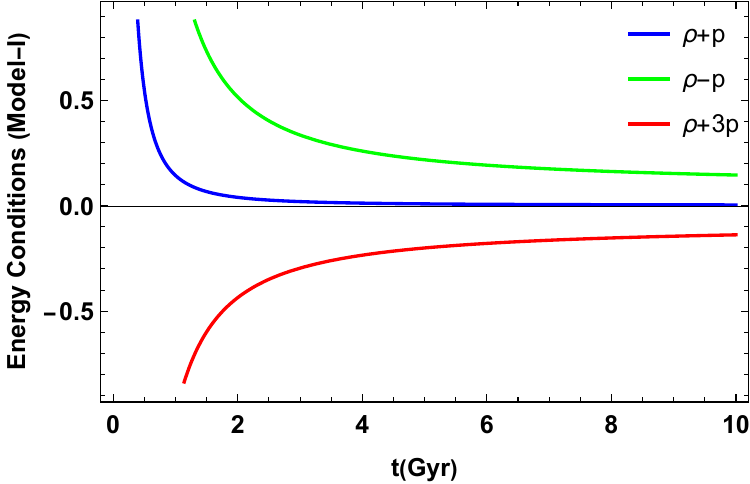}
\caption{Model I}
\label{VIII a}
\end{subfigure}
\hfill
\begin{subfigure}[b]{0.33\textwidth}
        \centering
        \includegraphics[width=\textwidth]{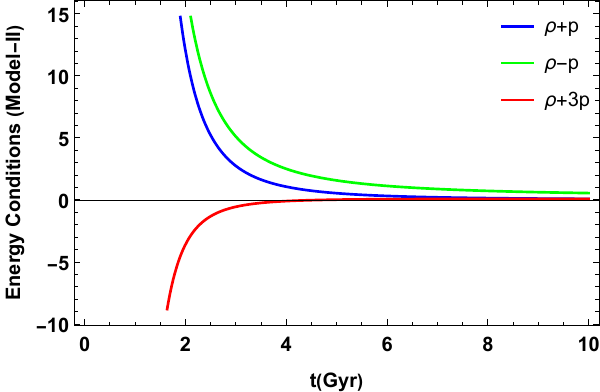}
\caption{Model II}
\label{VIII b}
    \end{subfigure}
    \hfill
\begin{subfigure}[b]{0.33\textwidth}    \centering
\includegraphics[width=\textwidth]{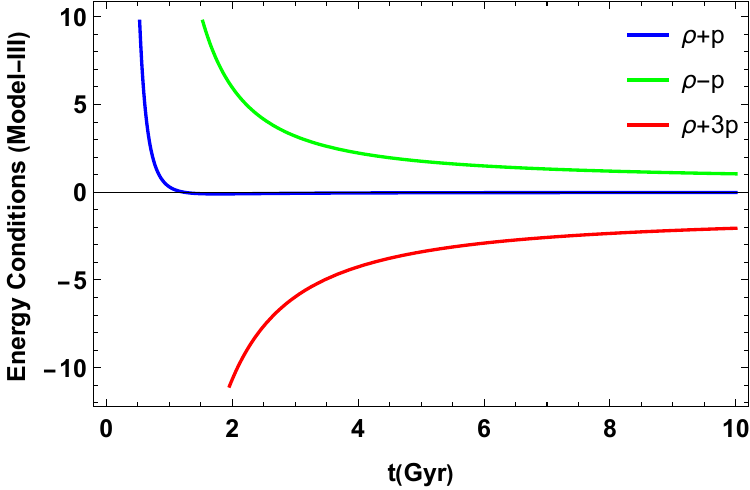 }
\caption{Model III}
\label{VIII c}
\end{subfigure}

\caption{These figures show the energy conditions at   $n=1.5$, $b=0.25$, $t_{0}=0.5$ and $k=0.8$ for Models- I, II and III, respectively.}
    \label{fig:VIII}
\end{figure}

\begin{equation*}
  1) \rho \geq 0, \rho \pm \Bar{p} \geq 0 (DEC)
     \quad 2) \rho \geq 0, \rho+\Bar{p}\geq0  (WEC) 
     \quad 3) \rho + \Bar{p} \geq 0, \rho+3\Bar{p}\geq 0(SEC)   
\end{equation*}

Figure \eqref{fig:VIII} shows that the weak energy condition (WEC) and dominant energy condition (DEC) are satisfied for the constructed model. However, the strong energy condition (SEC), is violated for each model. There is a reverse gravitational impact when the (SEC) is violated. The cosmos jerks as a result, and the current acceleration phase gives way to the previous deceleration phase \cite{caldwell2006sudden}. As a result, the model proposed in this research appears to be a good fit for explaining the universe's late temporal acceleration.

\section{conclusions}\label{VI}
Here, we have described the spatially homogeneous and anisotropic Kantowski-Sachs space-time in the presence of a bulk viscous fluid with one-dimensional cosmic strings in $f (T )$ gravity by modifying general relativity to explain the challenging problem of late time acceleration of the universe. To obtain a definite solution to the highly non-linear field Eqs. \eqref{17}$-$\eqref{19} of this theory, we use the special law of variation \eqref{26}$-$\eqref{28}.

Interestingly, the Hubble parameter $H$ decreases and ultimately approaches a small positive value within the passage of cosmic time. Figure \eqref{I a} shows that the model's expansion is faster at the beginning of the universe and subsequently slows down over time. This demonstrates that the universe's evolution begins at an infinite rate and decreases as it expands. The deceleration parameter $q$ rapidly decreases and asymptotically approaches $-1$, demonstrating a de Sitter-like expansion at a late time, which explains the acceleration of the universe at a late time. Figure \eqref{I b} displays that the universe shows the phase transition from current deceleration to late-time acceleration. Hence, for the late time, we get $H>0$ and $q<0$ implying that the models show the expanding and accelerating nature of the universe.
Figure \eqref{fig:II} represents the behavior of the total energy density $ \rho$. It is observed that $\rho$ is a positive decreasing function of time, which shows the expansion nature of the universe. The negative behavior of the bulk viscous pressure $\Bar{p}$ is shown in Fig. \eqref{fig:III} which corresponds to cosmic acceleration according to standard cosmology. Every model shows negative trajectories throughout cosmic evolution. Model I approaches a small negative value, while models II and II approach zero for a later time. Therefore, the constructed models shows an accelerated expansion phase in the future. One important parameter that characterizes many matter-dominated periods in the development of the universe is the EoS parameter $\omega_{eff}$. Figure \eqref{fig:IV}  depicts the behavior of the EoS parameter against cosmic time $t$. Through an analysis of model I, \eqref{IV a}, we found that the EoS parameter takes on the value $\omega_{eff} \sim - 1 $  during the early stage and then entered a quintessence in era for later times. This result is consistent with certain EoS constraints that have recently been linked to the inflationary EoS. As time $t$ progresses, model II  (Fig. \eqref{IV b}) behaves like the $\Lambda$CDM model. For model III, the effective EoS parameter $\omega_{eff}$ figure \eqref{IV c} resembles the dark energy model and behaves like the quintessence model $ - 1 < \omega_{eff} <0$ throughout the cosmic evolution. The string density $\lambda$ is a decreasing function of time and is always greater than zero for models I and II. 
Figures \eqref{V a} and \eqref{V b} show the plots of string energy density versus
time $t$ for models I and II. The $\lambda$ remains positive for both models.  However, it decreases more sharply
with increasing cosmic time. In the early phase of the universe, the string energy density of both models will dominate the
dynamics, and later, it approaches zero.
However, this becomes a reversible case for model III. The string energy density $ \lambda$ is negative throughout cosmic evolution. Figure \eqref{V c}  shows that, at late times, $\lambda$  approaches zero. Finally, we can conclude that for every model at the present epoch, we obtain a string-free model, and the string phase of the universe disappears, i.e., we have an anisotropic fluid of particles. 
The cosmic fluid dynamics of the universe are significantly affected by the positive decrease and ultimate decrease in the coefficient of bulk viscosity $\xi$ for model II \eqref{VI b}, but the coefficient converges to a small positive value for models I and III [Figs. \eqref{VI a} and \eqref{VI c}. This decrease in bulk viscosity might lead to a further acceleration of the universe, potentially causing it to expand at an even faster rate. As a result, the universe could enter a phase of rapid expansion, similar to inflationary models, which have been proposed to explain the observed accelerated expansion  Furthermore, the decrease in viscosity could help to prevent the initial singularities, consistent with models indicating the beginning of a smooth universe. Additionally, a decrease in bulk viscosity can enhance the role of dark energy or modified gravity in driving accelerated expansion. From Fig. \eqref{VII a} and \eqref{VII c}, we observe that for $k=1$, every model becomes isotropic and shear-free. From Fig. \eqref{fig:VIII}, we confirmed that for models I to III, the WECs and DECs are valid but violate the SEC. The violation of the SECs has an anti-gravitational effect for which the universe is jerked;  thus, our universe exhibits a transition from early decelerating to the present accelerating phase.

Understanding the behavior of the accelerating universe under the influence of teleparallel gravity can provide insights into cosmic evolution. The study of the transition from a decelerated to an accelerated phase in the universe can contribute to our knowledge of the dynamics of the cosmos. Investigating the three different models of $f(T)$ gravity which is a specific form of teleparallel gravity and its impact on the acceleration behavior of the universe may have implications for cosmological models and theories are the major applications of this constructed model.  Further exploration of the impact of the bulk viscousity content in various cosmological scenarios could enhance our understanding of cosmic evolution.  Investigating the behavior of bulk viscous pressure and pressure parameters across cosmic evolution may lead to insights into the mechanisms driving cosmic acceleration. Future studies could focus on refining the highly nonlinear form of teleparallel gravity used in the model to improve the accuracy of predictions regarding the accelerating universe. Every constructed model has some loopholes, and the model constraints may limit the generalizability of our findings. The highly nonlinear form of teleparallel gravity used in the model may introduce complexities that could affect the interpretability of the results.

\bibliography{YBI}
\bibliographystyle{rsc}
\end{document}